# Influence of the inner shell electrons on photoionization of outer shells

**M. Ya. Amusia[1;2], L. V. Chernysheva[2] and E. G. Drukarev[3]**

[1]*The Racah Institute of Physics, the Hebrew University of Jerusalem, Jerusalem 91904, Israel*
[2]*A. F. Ioffe Physical-Technical Institute, St. Petersburg 194021, Russian Federation*
[3]*National Research Center "Kurchatov Institute", B. P. Konstantinov Petersburg Nuclear Physics Institute, Gatchina, St. Petersburg 188300, Russian Federation*

**Abstract:**

We have studied the role of virtual excitations of inner shells upon outer shell photoionization. The calculations were performed in the frames of the Random Phase Approximation and its generalized version's GRPAE that take into account variation of the atomic field due to electron elimination and the inner vacancies decay. We employ both analytic approximation and numeric computations. The results are presented for 3*p* electrons in Ar and 4*d*-electrons in Pd near inner shells thresholds. The effect proved to be noticeable.

**1**. The role of intershell interaction or electron correlations was disclosed theoretically long ago [1] in the frame of the Random Phase Approximation with Exchange (RPAE) (see [2] for details). Theoretical predictions were confirmed experimentally very soon. That formed the bases for an entire domain in photoeffect studies. They have developed theoretically and experimentally very fast (see [2] and reference therein). At first, main attention was concentrated on studies of low photon energy photoionization, namely on elimination of electrons from outer, sub-valent and some intermediate shells. It was found also that electron correlations are important close to inner shell thresholds.

Increasing accuracy of photoionization studies achieved using the synchrotron radiation permitted to look carefully upon phenomena in a broad photon energy region. As a result, it has been found relatively recently that correlations are important also at high photon frequency [3, 4] if interacting electrons belong to subshells with close values of energies, when the photon energy is high for both of them. Recently strong interaction within two components of intrashell doublet was described [5].

At the same time, if one considers photoionization of an outer shell close to the inner threshold, the photon energy is high for the outer shell, being low for the inner shell. There essential correlation effects may be of importance due to near threshold jump of the inner shell photoionization cross-section. Qualitatively, it has been predicted long ago (see [2], Chap. 5).

In the present Letter we consider this effect using both analytic evaluations and numerical calculations that are well more accurate than old estimations in [2]. We present here the results of analytical and numerical calculations for the cases of Ar, where 1*s* and 2*p* electrons affect the photoionization of the $3p^6$ and $5s^2$ electrons and Pd, where 3*p* affects the outer $4d^{10}$ electrons [6]. The effect proved to be essential and relatively easy detectable.

**2.** We employ here the RPAE approximation and its generalized version GRPAE [2]. For inner shells in RPAE the diagrammatic expression for the photoionization amplitude *D* looks as follows:

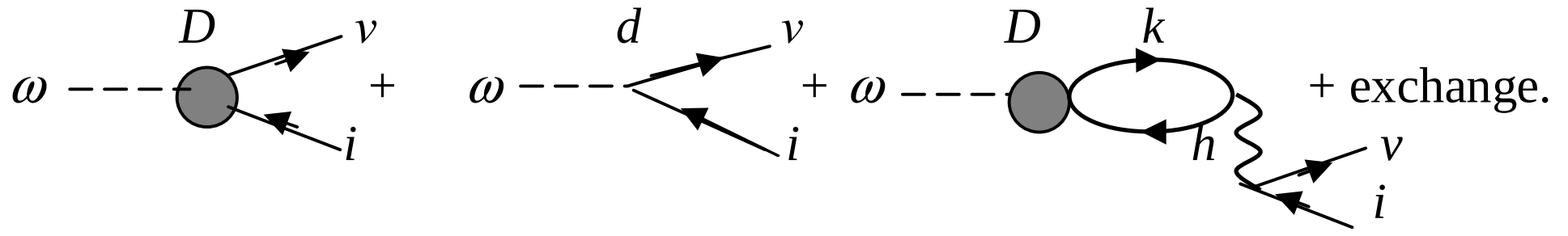

(1)

Here the gray circle denotes *D*, line with an arrow directed to the right (left) denotes an excited electron (vacancy), and the wavy line stands for the Coulomb interparticle interaction *V*. At high $\omega$ one can rightfully neglect the so-called time-reverse terms [2]. Analytically, (1) has the following form

$$\left\langle \nu \left| \hat{D}(\omega) \right| i \right\rangle = \left\langle \nu \left| d \right| i \right\rangle + \sum_{h \le F} \int_{k>F} \left\langle k \left| \hat{D}(\omega) \right| h \right\rangle \frac{\left\langle \nu h \left| V \right| ik - ki \right\rangle}{\omega + \varepsilon_h - \varepsilon_k - i\delta} \equiv d_{\nu i} + \Delta D_{\nu i} \tag{2}$$

Here $i$ and $\nu$ are the occupied states of the discrete spectrum and the state of the continuum, correspondingly. Also, $k > F\,(h \le F)$ are the vacant (occupied) states ($\varepsilon_h < 0$), the sum over $k$ includes integration over continuous spectrum; $\delta \to +0$. The vacant and occupied states are described in Hartree-Fock (HF) approximation for the neutral target atom, while in GRPAE the excited inner shell states are described by HF for an ion with a vacancy in this same shell [2]. As to the so-called time-reverse diagrams (see in [2]) that are omitted in (2), their contribution is small due to big virtuality of the corresponding intermediate state.

**3**. To clarify the origin of the phenomenon under discussion we limit the sum over occupied states *h* by one term $j$ and assume that excitations of this shell to the discrete spectrum can be neglected. Under this assumption we can put $\delta = \gamma_j / 2$, with $\gamma_j$ being the sum of the Auger and radiative widths of the hole $j$.

The integral over the energies of the continuum states $k$ is enhanced by the pole at $\omega = \varepsilon_k + \varepsilon_j + i\gamma_j / 2 \equiv \varepsilon_k - I_j + i\gamma_j / 2$, particularly close to inner shell threshold $\omega \simeq -\varepsilon_j \equiv I_j$, where $I_j$ is the *j*-shell ionization potential. The electron in the state $k$ is well screened and "feels" the effective charge $Z_{eff} > 1$. Putting $\varepsilon_k = 0$ in numerator of Eq.(2) we find that the integral over $k$ is saturated by energies $Z_{eff}^2 / 2 \gg \varepsilon_k > |\,\omega - I_j\,|$. In this region one can neglect the dependence of $\langle k \,|\, \hat{D}(\omega) \,|\, j \rangle$ and $\langle \nu j \,|\, V \,|\, ik - ki \rangle$ upon $\varepsilon_k$, and put $\varepsilon_k = 0$. Now Eq.(2) can be presented as

$$\Delta D_{\nu i} \approx \left\langle 0 \left| \hat{D}(I_j) \right| j \right\rangle \left\langle \nu j \left| V \right| i0 - 0i \right\rangle L_j(\omega - I_j) \equiv D_{0j}(I_j) U_{\nu j, i0} L_j(\omega - I_j), \tag{3}$$

where

$$L_j(\omega - I_j) \approx -\frac{1}{2} \ln \left\{ Z_{eff,j}^4 \Big/ \left[ 4\left(\omega - I_j\right)^2 + \gamma_j^2 \right] \right\}. \tag{4}$$

In (3) $0$ denotes the continuum state with $\varepsilon_k = 0$. To derive this formula, it was essential to take into account that the inner shell photoionization amplitude is a step-function that is zero below threshold, at $\omega < I_j$, and quite big already above threshold.

At $\omega$ close to $I_j$ the value $L_j$ becomes large, even diverges if the width $\gamma_j$ is neglected, thus essentially increasing the near-threshold amplitude. This explains qualitatively why RPAE predicts narrow spikes at photoionization threshold for e.g. 2p subshell in Ar (see [2]).

If it is necessary to take into account the discrete excitation levels of the inner shell $j$ it can be done substituting $D_{0j}(I_j)$ in (4) by $D_{0j}^{(1)}(I_j^{(1)}) = \left[ f_j^{(1)} I_j^{(1)} / 2\Delta I_j^{(1-2)} \right]^{1/2}$, where $f_j^{(1)}$ is the oscillator strength of the first exited level, $I_j^{(1)}$ is the first discrete excitation energy of the $j$ shell and $\Delta I_j^{(1-2)}$ is the energy distance between the first and second discrete excited level.

In order to estimate the value of $\Delta D_{\nu i}$, let us note that the energies of the outer electrons are much smaller than those of the inner ones, i.e. $|\varepsilon_j| \ll |\varepsilon_i|$. Thus, the average moments of the outer electrons $p_i = (2|\varepsilon_i|)^{1/2}$ are much smaller than those of the inner electrons $p_j = (2|\varepsilon_j|)^{1/2}$, i.e. $p_i \ll p_j$. Since we consider energies $\omega$ close to the ionization threshold of the inner electron, momentum of the photoelectron $p_\nu$ is big, $p_\nu \approx p_j$. Thus $p_i / p_j \ll 1$ is the only small parameter in our problem. Ionization of the outer electron at the considered photon energies takes place at the distances of the order $1/p_j \ll 1/p_i$.

The matrix element $D_{\nu i}$ is proportional to the radial function $R_i$ of the outer electron at small distances of the order $1/p_j$, i.e. $R_i(1/p_j)$ that is proportional to the small factor $\lambda_j = (p_i / p_j)^4 \ll 1$. In the matrix element

$$U_{\nu j,i0} \equiv \left\langle \nu j \middle| V \middle| i0 - 0i \right\rangle = \int d\mathbf{r}_1 d\mathbf{r}_2 \psi_j^*(\mathbf{r}_1)\psi_\nu^*(\mathbf{r}_2)\,|\,\mathbf{r}_1 - \mathbf{r}_2\,|^{-1} \left[\psi_i(\mathbf{r}_1)\psi_0(\mathbf{r}_2) - \psi_0(\mathbf{r}_1)\psi_i(\mathbf{r}_2)\right], \qquad (5)$$

which enters Eq.(3), the contributing distances *r* are the order of the inter bound electron size $r_j \sim 1/p_j$. The matrix element $\left\langle \nu j \middle| V \middle| i0 \right\rangle$ describes the mechanism, in which the electron moves from the state *i* to the hole in the state *j*, pushing the electron from the continuum state *k* to the continuum state with large momentum of the order $p_j$. It contains the small factor $R(p_i / p_j) \sim \lambda$.

The same refers to the exchange matrix element $\left\langle \nu j \middle| V \middle| 0i \right\rangle$. It describes the process, in which the electron moves from the continuum state *k* to the hole in the state *j*, pushing the electron from the state *i* to continuum. As one can easily demonstrate, it also contains the small factor $R(p_i / p_j) \sim \lambda_j$. Thus, we can estimate $\left\langle \nu j \middle| V \middle| i0 \right\rangle \sim \left\langle \nu j \middle| V \middle| 0i \right\rangle \sim \lambda p_j$.

The matrix element $\left\langle 0 \middle| \hat{D}(I_j) \middle| j \right\rangle$ is also proportional to $\lambda_j$. We have assumed earlier that the intermediate electron is described by the Coulomb field function with effective charge of the nucleus $Z_{eff,j}$. This leads to estimation

$$\frac{\Delta D_{\nu i}}{D_{\nu i}} \sim \frac{D_{0j}}{D_{\nu i}} L_j U_{\nu j,io} . \qquad (6)$$

Note the essential difference between the case, considered in [4], and this one (see also [6]). In the case of photon energies that are much higher than both energies of interacting closely located subshells [4], the transferred via Coulomb interaction linear momentum is big, of the order of $q \sim \sqrt{2\omega}$, leading to a small factor in the direct amplitude's contribution $(I_i / \omega)^2$. As to the exchange term in (1), for it the typical transferred momentum $q \sim \sqrt{2I_i}$ is small. So, in the total amplitude in (1) the exchange term dominates

The RPAE and GRPAE calculations include this contribution automatically. However, once we clarified the main mechanism of the phenomena, we can try to make crude estimations by employing the Coulomb functions with Slater effective charges [7].

Let us start with the influence of the 1s electrons on ionization of the 3p state in Ar. The effective nuclear charges for the electrons in 1s and 3p states are $Z_{1s} = 17.7$ and $Z_{3p} = 6.75$. Denote as $N_\nu$ the normalization factor of the continuum wave function with asymptotic momentum $\mathbf{p} \to 0$. For the partial wave expansion with the radial functions normalized to the delta function of the energy $N_\nu = \sqrt{Z_{eff} / c}$ with $Z_{eff}$ being the effective value of the nuclear charge felt by the electron in the continuum state with the energy $\varepsilon = 0$. We assume that it is the same as the one felt by the external bound electron, i.e. $Z_{eff} = Z_{3p}$. We find that $U_{\nu j,i0} = N_\nu U.$, while in the lowest order of expansion in powers of $Z_{3p}^2 / Z_{1s}^2 \ll 1$

$$U_{\nu j,i0} \equiv N_\nu U = -8\pi N_\nu \frac{\sqrt{2}}{27} \left( \frac{Z_{3p}}{Z_{1s}} \right)^{5/2} Z_{1s} . \qquad (7)$$

Thus, $U = -2.1$ and the correlations lead to increase the cross section since $L < 0$. We define $D = D_{01s} / (N_\nu D_{\nu 3p})$, and the product $N_\nu D$ is proportional to $(Z_{1s} / Z_{3p})^{5/2}$. In the final expression

$$\frac{\Delta D_{\nu 3p}}{D_{\nu 3p}} = \frac{UDLZ_{3p}}{2\pi c} \qquad (8)$$

the product $DU = 2e^{-2} Z_{1s}$. ($e^{-2} \approx 0.135$ is the characteristic factor of the Coulomb matrix element at the threshold) does not depend on the effective charge felt by the external 3p electron. Numerically, one has $\Delta D_{\nu 3p} / D_{\nu 3p} = 1.8$. Thus, correlations with the 1*s* electron effectively increase the cross section by a factor of about $3$.

The situation for correlations with 2s state in ionization of 3p state is more complicated since in the lowest order of $Z_{3p}^2 / Z_{2s}^2$ we find $U = 0$. The Coulomb calculation beyond the leading order in $Z_{3p}^2 / Z_{2s}^2$ provides $DU = 2e^{-2} (Z_{3p}^2 / Z_{2s}^2) Z_{2s}$. Numerically, one has $\Delta D_{\nu 3p} / D_{\nu 3p} = -0.48$.

Note, however that the Coulomb function for the state 2s contains a polynomial of the first order $1 - rZ_{eff}/2$, and any deviation from pure Coulomb shape $1 - arZ_{eff}/2$, where $a \neq 1$, provides a nonzero value in the lowest order of expansion in powers of $Z_{3p}^2/Z_{2s}^2$. Thus if we parametrize the wave function at the distances of the order of the size of 2s shell as $\psi = e^{-Z_{eff}r}(1 - arZ_{eff}r)/2$, we find a nonzero correlation effect in the lowest order in $Z_{3p}^2/Z_{2s}^2$ ($\Delta D_{v3p}/D_{v3p} > 0$ if $a$<0).

We consider also ionization of 4*d* state in Pd that is affected by the 3*d* electrons. Here the effect is of the opposite sign, i.e. the correlations diminish the value of the cross section. This becomes clear if we recall that the wave function of 3*d* state turns to zero only at the origin, and it has additional factor $r^2$ comparing to that of 1*s* state. Actually we must calculate the Fourier transform of the wave function of internal electron. Each power of *r* corresponds to the factor $ip$, and thus the factor $r^2$ changes the sign of the Fourier transform.

So, our analytic considerations predict a noticeable variation in the outer shell photoionization cross-section for photon energies close to inner shell ionization (excitation) threshold. This variation appears due to sharp increase of the inner shell ionization cross-section near ionization/excitation thresholds. Of course, the factor $L_j(\omega - I_j)$ in (3) even for very small $\gamma_j$ is small as compared to $p_j$. But due to very fast variation with $\omega$ it has to be visible as cross-section variation at $\omega$ close to the corresponding limiting values, below $\omega = I_j^{(1)}$ and above $\omega = I_j$. We predict also the change of sign of the effect for correlations with d states

**4.** The estimations and crude calculations performed above elucidate the nature of the considered effect, but are unable to present quantitatively reliable data. To obtain such data, we have solved the eq. (2) numerically for two atoms, Ar and Pd. We employ our RPAE computing codes that use as one-electron the Hartree-Fock (HF) approximation. The calculation details one can find in [8, 9].

It is known, that while RPAE works very well for outer and intermediate shell, it is not sufficient to describe photoionization of inner and innermost subshells and shells. In these cases RPAE has to be complemented by taking into account effects of so-called static rearrangement for inner shells and vacancy decay along with static relaxation for the innermost [9].

Static rearrangement takes into account the modification of the self-consistent HF field that acts upon excited or ionized, really or virtually, electron, due to modification of other atomic electron states caused by creation of the vacancy *j*. The creation of vacancy *j* affects all occupied states *h* that enter the sum over *h* in (2). Due to presence of the vacancy *j* all atomic electron feel additional attraction to the nucleus that increases screening of vacancy *j* field acting upon excited or ionized electron *k*.

Vacancy *j* decay, on the contrary, increases the strength of the HF field acting upon the outgoing or exited electron. Both effects, namely static screening and vacancy decay with good accuracy can be taken into account by calculating the electron *k* HF wave functions, using different options, namely the "frozen core" HF approximation, rearranged HF of an atom with the vacancy *j* and HF field in the state that is formed after vacancy *j* decay.

The results of calculating the *i* electron photoionization cross-sections obtained using the amplitude, given by (2) in three mentioned above one-electron approaches are denoted on figures that follows, as, respectively, RPAE, GRPAE and GRPAEII. The results without taking into account the influence of the *j* subshell are denoted as HF. The results are collected in Fig. 1 and 2.

In Fig. 1 we present the vicinity of two thresholds, 2s and 1s for photoionization cross-section of the outer 3*p* electrons in Ar. Similar data for 4*d* electrons in Pd near 3*d* and 3*s* thresholds are depicted in Fig. 2. It is seen that in all presented cases the role of the inner shell below and above its threshold is quite prominent and stretches well outside the discrete excitations photon energy region $|I_j - I_j^{(1)}|$, just as it was demonstrated by the analytic model in the previous section. Calculations show that correlational can be of different sign, leading to increase of the cross-section and its decrease, again in accord with the previous section analytic model. An essential influence of correlation effect beyond the RPAE scope was found. This is illustrated in Fig. 1, where results are presented not only in RPAE, but GRPAE and GRPAEII.

In fact, we have calculated much more cross-sections than presented in the Figures. However, we did not found an example of opposite trends at photon energies below and above the discrete excitations region. Note that such a trend is not excluded by analytic considerations.

**5.** The results presented here, along with previous calculations of 3*d* photoionization cross-section near 2*p* threshold in Pd and experimental results on the same subshells in In [6], is a new area in photoionization studies that is far from being satisfactorily explored. In this area the intershell interaction is visible but relatively small, so it can be accounted for perturbatively, thus verifying the ability of the approaches employed to describe both, the photoionization amplitude far from its threshold and the intershell interaction.

One of us (EGD) acknowledges the support by the grant RSCF 14-22-00281.

**Figures** (color on line)

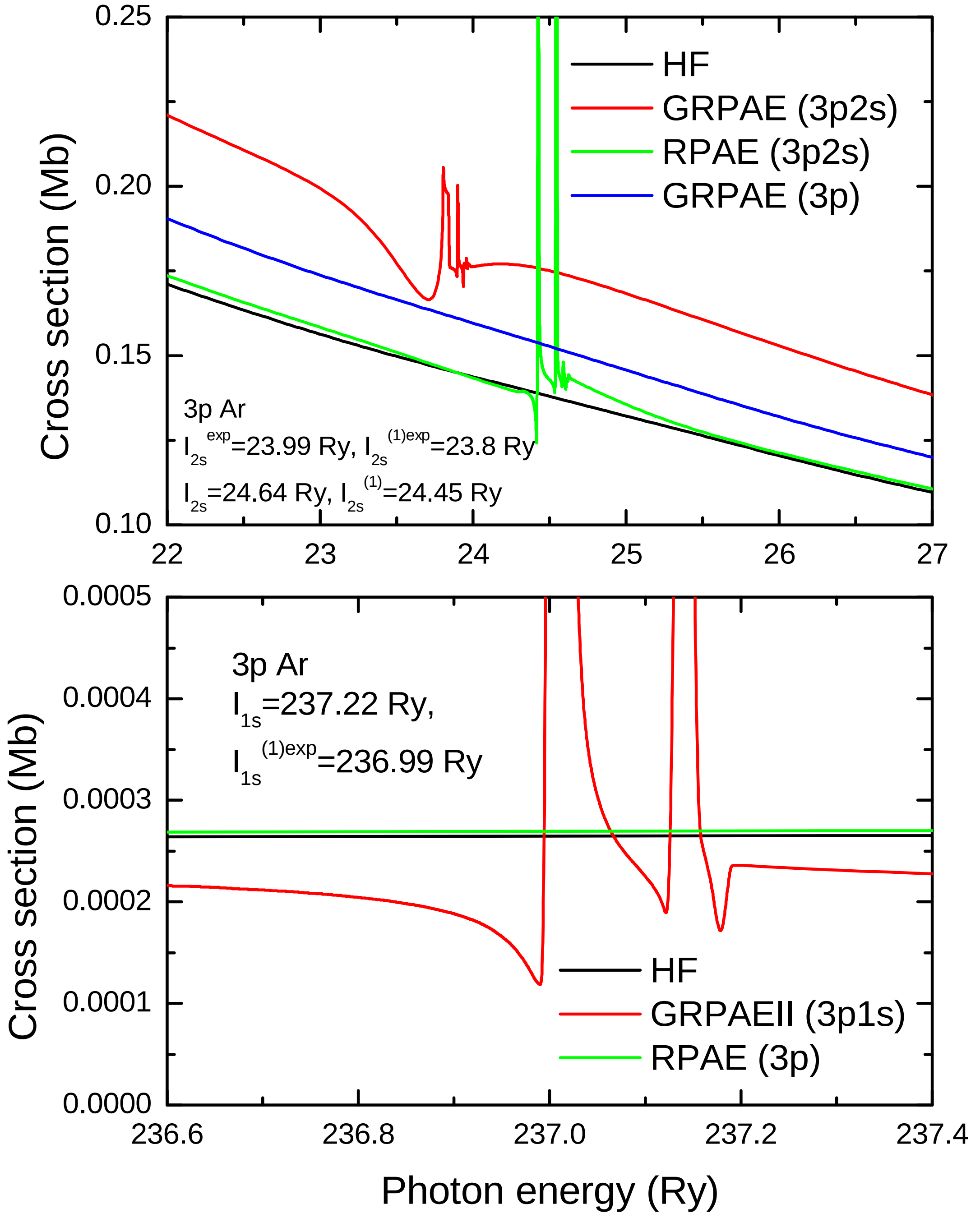

Fig. 1. Photoionization cross-section of 3p electrons near 1s and 2s thresholds of Ar, respectively.

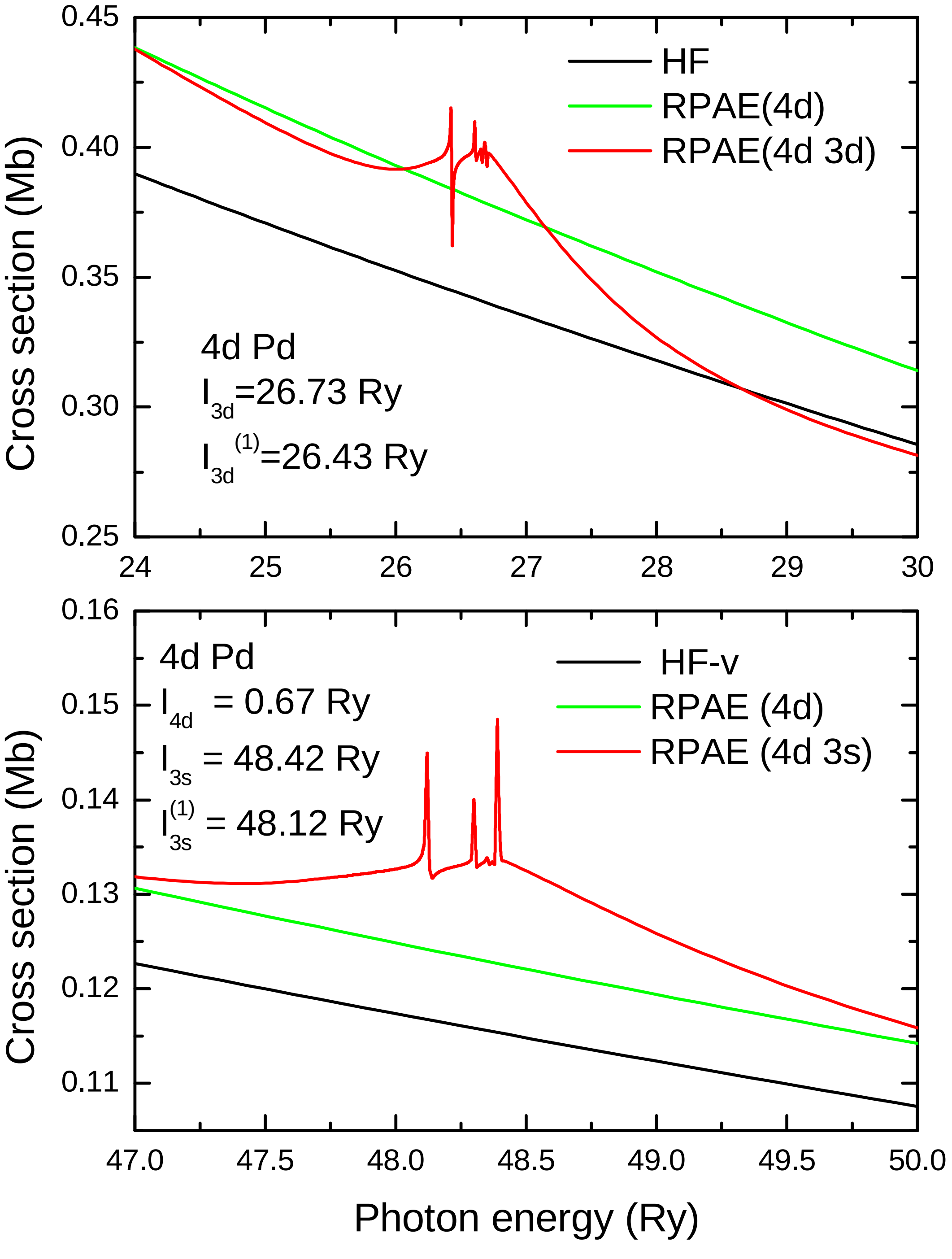

Fig. 2. Photoionization cross-section of 4d electrons near 3d and 3s thresholds of Pd, respectively.